\documentclass[12pt]{article}
\usepackage{graphics}
\usepackage{amssymb}

\begin{document}

\parskip=2pt
\parindent=7mm
\renewcommand{\baselinestretch}{1.}
\renewcommand{\theequation}{\arabic{section}.\arabic{equation}}

\newcommand{\be}{\begin{equation}}
\newcommand{\ee}{\end{equation}}
\newcommand{\ba}{\begin{eqnarray}}
\newcommand{\ea}{\end{eqnarray}}
\newcommand{\bd}{\begin{description}}
\newcommand{\ed}{\end{description}}
\newcommand{\rhs}{{\em rhs }}
\newcommand{\lhs}{{\em lhs }}
\newcommand{\pd}{\partial}
\newcommand{\e}{{\rm e}}
\newcommand{\D}{{\rm d}}
\newcommand{\C}{\mathbb{C}}
\newcommand{\N}{\mathbb{N}}
\newcommand{\R}{\mathbb{R}}
\newcommand{\SSS}{{\mathbb{S}}}
\newcommand{\Z}{\mathbb{Z}}
\newcommand{\DD}{{\mathcal{D}}}
\newcommand{\OO}{{\mathcal{O}}}
\newcommand{\UU}{{\mathcal{U}}}
\newcommand{\EE}{{\mathcal{E}}}
\newcommand{\FF}{{\mathcal{F}}}
\newcommand{\LL}{{\mathcal{L}}}
\newcommand{\QED}{\mbox{\rule[-1.5pt]{6pt}{10pt}}}
\newtheorem{claim}{Claim}[section]
\newtheorem{theorem}[claim]{Theorem}
\newtheorem{proposition}[claim]{Proposition}
\newtheorem{corollary}[claim]{Corollary}
\newtheorem{lemma}[claim]{Lemma}
\newtheorem{remark}[claim]{Remark}
\newtheorem{remarks}[claim]{Remarks}
\newcommand{\eg}{{\em e.g.}}
\newcommand{\ie}{{\em i.e.}}
\newcommand{\cf}{{\em cf. }}

\title{Berry phase in magnetic systems \\ with point perturbations}
\date{}
\author{Pavel Exner$^{a,b}\!$ and Vladimir A. Geyler$^{c}\!$}
\maketitle

\begin{quote}
{\small {\em a) Nuclear Physics Institute, Academy of Sciences,
25068 \v Re\v z \\ \phantom{a) }near Prague, Czechia
 \\ b) Doppler Institute, Czech Technical
University, B\v rehov{\'a} 7,\\ \phantom{a) }11519 Prague,
Czechia}
 \\ {\em c) Department of Mathematical Analysis, Mordovian State \\
 \phantom{a) } University,  430000 Saransk, Russia;}\\
 \phantom{a) }\texttt{exner@ujf.cas.cz},
 \texttt{geyler@mrsu.ru} }
\end{quote}

\begin{abstract}
\noindent We study a two-dimensional charged particle interacting
with a magnetic field, in general non-homogeneous, perpendicular
to the plane, a confining potential, and a point interaction. If
the latter moves adiabatically along a loop the state
corresponding to an isolated eigenvalue acquires a Berry phase. We
derive an expression for it and evaluate it in several examples
such as a homogeneous field, a magnetic whisker, a particle
confined at a ring or in quantum dots, a parabolic and a
zero-range one. We also discuss the behavior of the lowest Landau
level in this setting obtaining an explicit example of the
Wilczek--Zee phase for an infinitely degenerated eigenvalue.
\end{abstract}


\section{Introduction}

A nontrivial Berry phase \cite{Ber} can be demonstrated in
different situations. There is a growing interest recently to this
effect in mesoscopic systems -- see, \eg, \cite{LSG,MHK} and
references therein. These papers investigate theoretically and
experimentally how the phase is manifested in quantum dynamics of
a particle with spin interacting with a time-dependent magnetic
field. In the present paper we are going to discuss a simple model
in which the Berry phase emerges even if the spin-orbital coupling
is neglected.

The model describes a charged particle confined to a potential
well and placed into a magnetic field of constant direction, which
is independent of time and may be homogeneous. The phase will
appear when the well is moving adiabatically. The similar situation
appears in the Born--Oppenheimer approach for the study of molecules
(see, \eg, \cite{Ja} and references therein) and impurities in
semiconductors \cite{Za2}. For the sake of
simplicity we suppose that the well represents a zero-range
interaction, \ie, it is given by a point interaction in the plane.
This makes it possible to derive explicit formulae for the Berry
potential. The idea of employing point interactions to this
purpose is not new: some solvable models exhibiting a nontrivial
Berry phase have been constructed earlier. For instance, the
geometric phase resulting from a cyclical motion of the boundary
condition for the Dirac and Schr\"odinger equations on an interval
$[0,\ell]$ was computed in \cite{BFG,GK}. On the other hand, the
authors of \cite{CS} investigated the Berry phase which arises
when a pointlike scatterer is adiabatically moved in a rectangular
billiard in such a way that the energy levels encircle a
``diabolic point".

In our cases the results are simpler and rather illustrative. In
particular, we shall show that moving the zero-radius potential
well along a closed curve in the plane, the eigenfunctions of a
particle trapped by the well and exposed to a homogeneous field
perpendicular to the plane acquire a phase which coincides with
the number of magnetic field quanta through the area restricted by
the curve. This picture changes if an additional confining
potential is added, say, in the form of an annular potential
``ditch". In the limiting case of an infinitely thin ring the
motion of the point interaction induces a geometric phase which
differs from the above one on a quantity proportional to the
persistent current in the annulus. Recall that persistent currents
in a ring with a point perturbation were investigated -- see, \eg,
\cite{CGR} -- but the relation to the Berry phase was not noticed
.

Let us describe briefly the contents of the paper. In the next
section we shall recall briefly how the zero-range interaction in
a magnetic system is constructed and how its spectrum is
determined by means of the Krein's formula. For simplicity we
suppose always that the magnetic field as well as the possible
confining scalar potential are rotationally symmetric. The central
part of the paper is Section~3 where we derive a general
expression for the Berry potential corresponding to a point
interaction moving along a smooth curve -- \cf Eq.~(\ref{angular
F}).

This result is in the next section illustrated on the number of
examples. We show that the Berry phase for the perturbation moving
along a closed loop $C$ in a homogeneous field without a scalar
potential is proportional to the number of flux quanta through
$C$. In distinction to that, the phase corresponding to a magnetic
whisker contains an extra term proportional to the persistent
current in the loop $C$. For comparison we analyze an electron
confined to a circular ring and find the same Berry phase
expression containing the persistent-current term, in this case
independently of the field profile. Finally, we discuss a harmonic
quantum dot in a homogeneous field. We show that if the point
interaction is strong the effect of the confining potential is
small and the Berry phase is again given by the number of the flux
quanta through $C$, up to an error term. We compare this with the
situation where the quantum dot itself is zero-range.

The behaviour of degenerate eigenvalues under adiabatic change of
parameters is more complicated and less understood. In the final
section address this question in the present setting and discuss
what happens in this situation with the lowest Landau level. We
compute the generalized Berry potential which determines the
corresponding Wilczek-Zee phase, and find the latter for adiabatic
evolution around a small loop. It appears to be nontrivial for the
angular momentum $m=1$ while the states with higher momenta are
not affected. Moreover, the phase which arises here differs in
sign from the one corresponding to the isolated energy level; we
explain this effect as a sort of topological charge conservation.


\setcounter{equation}{0} \section{Magnetic systems with a point \\
perturbation}

As indicated above we shall consider a charged particle of charge
$e$ and mass $m_*$ (which may be thought of as the effective mass
of an electron in a crystal) living in the plane with Cartesian
coordinates $x,y$ and exposed to a magnetic field perpendicular to
the plane, $\vec B= B(x,y) \vec e_z$. We also assume that the
particle may be confined to a part of the plane by a non-negative
potential $W$. The main simplifying assumption we shall make
concerns the  {\em rotational symmetry:} we suppose that there is
a system of polar coordinates $r,\varphi$ such that the magnetic
field and the confining potential depend on the radial coordinate
only, $B= B(r)$ and $W= W(r)$. In this case one can choose a gauge
in such a way that the radial component of the vector potential
vanishes, $A_r(r,\varphi)=0$, and $A_{\varphi}(r,\varphi)=
A_{\varphi}(r)$ depends on $r$ only. In particular, $\nabla \vec
A=0$.

It is convenient to single out the uniform component of the
magnetic field, $\vec B= \vec B_0+ \vec B_1$ with $\vec B_0$ being
a fixed vector. Of course, such a decomposition is arbitrary, but
we will have mostly in mind situations when $\vec B$ has a finite
limit as $r\to\infty$; then the non-uniqueness is removed by the
requirement $\vec B_1\to 0$. We shall also employ the
corresponding decomposition of the vector potential, $\vec A= \vec
A_0+ \vec A_1$. In view of the assumed symmetry it is natural to
use the circular gauge, $\vec A_0(r)= {1\over 2} B_0r\, \vec
e_{\varphi}$. As for the nonconstant part, we are particularly
interested in the example of an infinitely thin Aharonov-Bohm
solenoid, or a magnetic flux line with $A_{1\varphi}(r)=
{\Phi \over 2\pi r}$, where $\Phi$ is the magnetic flux through
the solenoid. It is convenient to use a dimensionless parameter
$\eta$, $\eta=\Phi/\Phi_0$, where
$$
\Phi_0= {2\pi\hbar c\over|e|}
$$
is the magnetic flux quantum; so $\eta$ is the number of quanta
carried by the solenoid. The corresponding magnetic field is
concentrated at the origin of the coordinates, $\vec B_1=
\Phi \delta(r) \vec e_z$.

In the following considerations, however, we assume only that
$A_{1\varphi}$ is a smooth function of the $r$ variable on the
halfline $(0,\infty)$. Under the stated assumptions, the particle
Hamiltonian has the following form:
\ba
\label{Ham2}
H &\!=\!& {1\over 2m_*} \left( -i\hbar\vec\nabla -\,{e\over
c}\vec A \right)^2 + W  \nonumber \\ &\!=\!& -{\hbar^2\over 2m_*}
\left\lbrack {1\over r} {\pd\over\pd r} r {\pd\over\pd r} +
{1\over r^2} {\pd^2\over\pd\varphi^2} \right\rbrack +
{e\hbar\over 2m_*c} \left( B_0 +\, {2\over r} A_{1\varphi}(r)
\right) \left(i{\pd\over\pd\varphi}\right)
\nonumber \\ && + {e^2\over 2m_*c^2}
\left( {1\over 4} B_0^2r^2 + B_0 r A_{1\varphi}(r) +
A_{1\varphi}(r)^2 \right) + W(r)\,. \label{Ham1}
\ea
The potential $W$ is non-negative by assumption, and therefore $H$
is a well defined self-adjoint operator, which can be understood,
\eg, as the Friedrichs extension of the operator defined on
$C_0^{\infty}(\R^2\setminus\{0\})$ by the \rhs of
Eq.~(\ref{Ham2}).

Now we shall introduce a point perturbation of the above
Hamiltonian located at a point $\vec s\in\R^2$ with the polar
coordinates $(\rho,\theta)$.
(Further we will assume that $\vec s\ne0$ if $\vec A_1$ has a singularity
at the point $r=0$).
The perturbed operator
$H_{\alpha,\vec s}$ is conventionally obtained as a self-adjoint
extension of the symmetric operator $S$ which is a restriction of
$H$ to the domain
\be
\label{restrdom}
\DD:= \{ \psi\in\DD(H):\: \psi(\vec s\,)=0\,
\}\,;
\ee
since the deficiency indices of $S$ are $(1,1)$ the extensions are
characterized by a single parameter $\alpha$. Under rather general
assumptions about the Hamiltonian \cite{GMC} the Green function
$G_{\alpha,\vec s}(\vec r,\vec r\,';E)$ of $H_{\alpha,\vec s}$ is
given by the Krein formula
\be
\label{krein}
G_{\alpha,\vec s}(\vec r,\vec r\,';E) = G(\vec
r,\vec r\,';E) - [Q(E;\vec s\,)+\alpha]^{-1} G(\vec r,\vec s;E)G(\vec
s,\vec r\,';E)\,,
\ee
where $Q(E;\vec s)$ is the so-called Krein ${\cal Q}$-function or
renormalized Green function at the diagonal point $(\vec s,\vec s\,)$,
\be
\label{kreinf}
Q(E;\vec s) := \lim_{\vec r\to \vec s}
\left\lbrack G(\vec r,\vec s;E) - {m_*\over \pi\hbar^2}
\ln |\vec r-\vec s\,|^{-1} \right\rbrack\,,
\ee
and $\alpha$ is the mentioned parameter. The latter is related to
the scattering length $\lambda$ of the point interaction by the
formula
\be
\label{scatlen}
\alpha = {m_*\over \pi\hbar^2} \ln\lambda^{-1}\,.
\ee
Less formally, the point perturbation at a point $\vec s$ may be
defined via the Fermi pseudopotential of the form
\be
\label{pseudo}
\mu\delta(\vec r -\vec s\,)\left(1-\ln|\vec r -\vec s\,|
(\vec r -\vec s\,)\nabla_{\vec r}\right),
\ee
where the coupling constant $\mu$ is related to the parameter
$\alpha$ by  $\mu=\alpha^{-1}$.

Under rather weak regularity requirements on the potentials
$\vec A(r)$ and $W(r)$ the Green function is of the form
\be
\label{greensing}
G(\vec r,\vec r\,';E) = {m_*\over \pi\hbar^2}
\ln |\vec r-\vec r\,'|^{-1} + G_0(\vec r,\vec r\,';E)\,,
\ee
where $G_0$ is continuous in the variables $\vec r, \vec r\,'$ and
analytic with respect to $E$ in the resolvent set, $\C\setminus
\sigma(H)$, of the free operator. It is the case, for example,
if $\vec A_1$ and $W$ are smooth functions (see, e.g., \cite{Be},
Chapter III, Theorem 5.1.). If $\vec A_1$ has a singularity at the
origin, then every point $\vec r$, $\vec r\ne 0$, has a neighborhood
such that Eq.~(\ref{greensing}) is true for $\vec r\,'$ in this neighborhood.

Since the singular term in Eq.~(\ref{greensing}) is energy
independent, $\pd G/\pd E$ needs no renormalization and we have
\be
\label{kreinder}
{\pd Q(E;\vec s\,)\over\pd E} = {\pd G\over\pd
E}(\vec s,\vec s;E)\,.
\ee
Due to the well-known Weyl theorem, the essential spectra of $H$
and $H_{\alpha,\vec s}$ coincide. As for the discrete spectrum, it
may happen that $H$ and $H_{\alpha,\vec s}$ have a common
eigenvalue. Let $E$ be an isolated eigenvalue of $H$ such that
there exists a corresponding eigenfunction $\psi$ satisfying
$\psi(\vec s\,)=0$ (in particular, this can be always achieved  if
$E$ is a degenerate eigenvalue). Then $E$ belongs to the spectrum
of $H_{\alpha,\vec s}$ as an isolated eigenvalue; moreover, the
multiplicity $m'$ of $E$ in the spectrum of $H_{\alpha,\vec s}$
obeys the inequality $m'\ge m-1$, where $m$ is the multiplicity of
$E$ in the spectrum of $H$. This assertion may be proven following
the arguments from \cite{CdV} where a special case of our claim
has been considered. In addition, the spectrum of $H_{\alpha,\vec
s}$ contains all solutions of the equation
\be
\label{energycon}
Q(E;\vec s\,)+\alpha = 0
\ee
(the true levels of the zero-range well). Every solution of this equation
lies in a gap of the unperturbed spectrum and is a simple isolated
eigenvalue of $H_{\alpha,\vec s}$. The corresponding eigenfunction $\psi$
has the form
\be \label{eif}
\psi(\vec r) = \left\lbrack {\pd Q\over \pd E}(E,\vec s)
\right\rbrack^{-1/2} G(\vec r,\vec s; E)\,.
\ee
Recall that in the real part of the resolvent set the derivative
is positive, $(\pd Q/\pd E)(E)>0$ for $E\in\R\setminus \sigma(H)$
-- \cf\cite{KL}. Therefore, equation (\ref{eif}) has at most one
solution in every gap of the spectrum $\sigma(H)$. Generally
speaking, the equation (\ref{energycon}) may have no solutions --
see, \eg, \cite{AGM}. It is
straightforward to see that if $E_0$ is an
isolated eigenvalue of $H$ and $\psi(\vec s\,)\ne 0$ holds for at
least one eigenfunction corresponding to $E_0$, then $E_0$ is a
pole of the function $Q(\cdot;\vec s\,)$. Hence if $\sigma(H)$ is purely
discrete solutions of Eq.~(\ref{eif}) exist in
infinitely many spectral gaps.

A simple but important particular case of the considered problem,
$\vec A_1=0$ and $W=0$, concerns a free motion in a uniform
magnetic field. In this situation the Green function acquires
the following explicit form,
\ba
\label{hom G}
G(\vec r,\vec r\,';E) &\!=\!& {m_*\over
2\pi\hbar^2}\, \Gamma\left( {1\over 2} - {E\over\hbar\omega_c}
\right) \exp \left\lbrack -\pi i\xi_0 \vec r\wedge\vec r\,' - {(\vec
r- \vec r\,')^2\over 4a_0^2} \right\rbrack \nonumber \\ &&
\times\Psi\left( {1\over 2}- {E\over\hbar\omega_c},\, 1; {(\vec r-
\vec r\,')^2\over 4a_0^2} \right)\,,
\ea
where $\omega_c$,
$$
\omega_c := {|eB_0|\over m_*c}\,,
$$
is the cyclotronic frequency, $\xi_0$,
$$
\xi_0 := {eB_0\over 2\pi\hbar c}\,,
$$
is the flux density of the uniform component of the
magnetic field, $a_0$,
\be
\label{magl}
a_0:= \sqrt{\hbar\over m_*\omega_c}= (2\pi|\xi_0|)^{-1/2}
\ee
is the magnetic length, and $\Psi$ is the Tricomi confluent
hypergeometric function -- \cf\cite{DMM}. The ${\cal Q}$-function
now obviously does not depend on $\vec s$ and equals
\cite{GM,GHS}:
\be
\label{hom Q} Q(E) = -\,{m_*\over 2\pi\hbar^2}\, \left\lbrack
\psi\left( {1\over 2} - {E\over\hbar\omega_c} \right) +2\gamma
-\ln2 -2\ln a_0 \right\rbrack\,,
\ee
where $\psi(x)= (\ln\Gamma(x))'$ and $\gamma= -\psi(1)$ is the
Euler constant. Up to a scaling and a shift in the argument the
behaviour of $Q(E)$ is given by that of the digamma function $\psi$; this
shows that in a uniform magnetic field the zero-range
potential with any fixed $\alpha\in\R$ induces existence of an
energy level on the halfline $(-\infty,\varepsilon_0)$ as well as
in each interval $(\varepsilon_\ell,\varepsilon_{\ell+1})$, where
$\varepsilon_\ell:= \left(\ell+{1\over 2}\right)\hbar\omega_c$ are
the Landau levels.


\setcounter{equation}{0} \section{The Berry phase}

Let us return to the condition (\ref{energycon}).
In what follows we will keep $\alpha$ fixed (and drop it mostly
from the notation) and move the point $\vec s$ along a smooth path
$C:\, \vec s=\vec s\,(t), \,t\in[0,1]$, in the plane $\R^2$
(or in the punctured plane $\R^2\setminus\{\vec 0\}$ if $\vec A_1$ has a
singularity at the point $\vec 0$) in such a way that Eq.~(\ref{energycon})
has a solution $E_0(\vec s,\alpha)$ lying in a gap of the unperturbed
Hamiltonian $H$. Denote
\be
\label{ef} \psi_{\vec s}(\vec r\,) = \left\lbrack {\pd Q\over
\pd E}(E_0(\vec s,\alpha),\vec s\,) \right\rbrack^{-1/2} G(\vec
r,\vec s; E_0(\vec s,\alpha))\,.
\ee
the corresponding normalized eigenfunction of
the perturbed operator $H_{\alpha,\vec s}$ (see (\ref{eif}).
If the path $C$ is a closed loop, $\vec s\,(0)= \vec s\,(1)$, the initial
and final state, $\psi_{\vec s(0)}$ and $\psi_{\vec s(1)}$,
respectively, differ by a phase factor,
\be
\label{loop phase}
\psi_{\vec s(1)} = \psi_{\vec s(0)} \exp
\left( -{i\over\hbar} \int_0^1 E_0(\vec s\,(t))\, \D t +i\gamma(C)
\right)\,, \ee
where the Berry phase $\gamma(C)$ depends only on the path $C$; in
accordance with Ref.~\cite{Ber} it equals
\be
\label{berry} \gamma(C) = \int_C \vec V(\vec s\,)\,\D\vec s
\,,
\ee
where
\be
\label{berry pot}
\vec V(\vec s\,) := i\langle \psi_{\vec s}\,|\,
\vec \nabla_{\vec s}\,|\, \psi_{\vec s} \rangle \,,
\ee
is the so-called Berry vector potential. Recall that from the
differential-geometric point of view ${\rm Im}\,\langle\psi_{\vec
s}\,|\,\vec\nabla_{\vec s}\,|\,\psi_{\vec s} \rangle$ is a
connection 1-form in a principal fiber bundle over $\R^2$ (or
$\R^2\setminus\{\vec 0\}$) associated with the eigenfunction
fibration $\psi_{\vec s} \mapsto \vec s$ \cite{Si}; in other
words, this quantity is a gauge potential with the gauge group
${\bf U}(1)$. We shall express $\vec V(\vec s\,)$ in the polar
coordinates,
\be
\label{pol berry}
\vec V(\vec s\,) = V_{\rho}(\rho,\theta) \vec
e_{\rho} + V_{\theta}(\rho,\theta) \vec e_{\theta}
\ee
with
\be
\label{comp berry}
V_{\rho} = i\langle \psi_{\vec s}\,|\,
\nabla_{\rho}\,|\,\psi_{\vec s} \rangle\,, \quad V_{\theta} =
{i\over\rho} \langle \psi_{\vec s}\,|\, \nabla_{\theta}\,|\,\psi_{\vec s}
\rangle\,.
\ee
To proceed further we need more information about the structure of
the Green function $G(\vec r,\vec r\,';E)$. First of all, we
decompose the state space $L^2(\R^2)$ into partial waves, \ie, we
represent it as $L^2(\R_+,r\,\D r)\otimes L^2(\SSS^1,\D\varphi)$
and perform the Fourier transform on the second component,
$L^2(\SSS^1,\D\varphi) \to \ell^2(\Z)$ with
\be
\label{four}
g \mapsto \{g_m\}_{m\in\Z}\,, \quad g_m =
{1\over\sqrt{2\pi}} \int_0^{2\pi} g(\varphi)\, e^{-im\varphi}
\D\varphi\,.
\ee
Then $L^2(\R^2)$ decomposes into an orthogonal sum of subspaces
each of which is isomorphic to the radial component,
\be
\label{orth dec}
L^2(\R^2) \simeq
\bigoplus_{m=-\infty}^{\infty} L^2(\R_+,r\,\D r)\,.
\ee
The unperturbed operator $H$ commutes with rotations around the origin, and
therefore it decomposes correspondingly into the orthogonal sum
\be
\label{H dec}
H = \bigoplus_{m=-\infty}^{\infty} H_m\,,
\ee
where the partial-wave parts $H_m$ are self-adjoint operators in
$L^2(\R_+,r\,\D r)$ obtained as the Friedrichs extensions of the
operators (\ref{Ham2}) with the domain $C_0^{\infty} (\R_+,r\,\D
r)$ and $-i\pd/\pd\varphi$ replaced by $m$. It is obvious that
each $H_m$ is a real operator, \ie, that it commutes with the
operator of complex conjugation in $L^2(\R_+,r\,\D r)$. It follows
that its Green function $G_m(r,r\,';E)$ is real valued for a real $E$.

The full Green functions can be expressed through its
partial-wave components as
\be
\label{Green}
G(\vec r,\vec r\,';E) = {1\over 2\pi}
\sum_{m=-\infty}^{\infty} e^{im(\varphi-\varphi')}G_m(r,r\,';E)\,.
\ee
It follows that $\langle \psi_{\vec s}\,|\, \nabla_{\rho}\,|\,\psi_{\vec
s} \rangle$ is a real number. On the other hand, differentiating the
identity $\langle \psi_{\vec s}\,|\,\psi_{\vec s} \rangle = 1$ we see that
the real part of $\langle \psi_{\vec s}\,|\, \nabla_{\rho}\,|\,\psi_{\vec s}
\rangle$ (as well as $\langle \psi_{\vec s}\,|\, \nabla_{\theta}\,|\,
\psi_{\vec s} \rangle$) vanishes. Consequently,
the radial component of the Berry potential $V_{\rho}(\vec s\,)=0$.
To find the angular one, let us differentiate the identity
\be
\label{Green id}
(H\!-\!E)G(\vec r,\vec s;E) = \delta(\vec r-\vec s\,)
\ee
with respect to the constant component $B_0$ of the magnetic field
keeping $E$ and $\vec s$ fixed; this yields
\be
\label{Green diff}
{\pd H\over\pd B_0}\,G + (H\!-\!E){\pd
G\over\pd B_0} = 0 \,.
\ee
Notice that ${\pd G/\pd B_0}$ is a smooth function in view of
(\ref{greensing}). Hence
$$
\left\langle G\, \Bigg|\, (H\!-\!E)\Bigg|\,{\pd G\over\pd B_0}
\right\rangle =
$$
$$
\left\langle \delta(\vec r-\vec s\,) \Bigg|\, {\pd
G(\vec r,\vec s;E) \over\pd B_0} \right\rangle = {\pd G\over\pd
B_0}(\vec s,\vec s;E) = {\pd Q\over\pd B_0}(E,\vec s\,)\,,
$$
and therefore
\be
\label{diff Q}
\left\langle G\, \Bigg|\, {\pd H\over\pd B_0}\,\Bigg|\,G
\right\rangle + {\pd Q\over\pd B_0} = 0\,. \ee
Dividing both terms of this expression by ${\pd Q/\pd E}$ and
putting $E=E_0(\vec s\,)$, we arrive at the relation
\be
\label{ident}
\left\langle \psi_{\vec s}\, \Bigg|\, {\pd
H\over\pd B_0}\,\Bigg|\,\psi_{\vec s} \right\rangle + {\pd Q\over\pd B_0}
\left( {\pd Q\over\pd E} \right)^{-1} = 0\,.
\ee
Since $E_0(\vec s\,)$ solves the equation
(\ref{energycon}), the last term at the \lhs can be expressed as
$$
\left.{\pd Q\over\pd B_0} \left( {\pd Q\over\pd E}
\right)^{-1}\right|_{E=E_0}= -{\pd E_0\over\pd B_0}\,,
$$
so
\be
\label{ident2}
\left\langle \psi_{\vec s}\, \Bigg|\, {\pd
H\over\pd B_0}\,\Bigg|\,\psi_{\vec s} \right\rangle = {\pd E_0\over\pd
B_0}\,.
\ee
Now we shall employ the formula (\ref{Ham1}) which yields
\be
\label{deriv} {\pd H\over\pd B_0} = i{e\hbar\over 2m_*c}\,
{\pd\over\pd\varphi} + {e^2B_0\over 4m_*c^2}\, r^2 + {e^2\over
2m_*c^2}\, r A_{1\varphi}(r) \,.
\ee
It follows from (\ref{Green}) that ${\pd\over\pd\theta} G(\vec
r,\vec s;E) = -{\pd\over\pd\varphi} G(\vec r,\vec s;E)$, and
furthermore, that $Q(E,\vec s\,)$ is independent of the angular
variable, $Q(E,\vec s\,)= Q(E,\rho)$. The last claim means that
$E_0(\vec s\,)$ also does not depend on $\theta$.
As a result we have that $\nabla_{\theta}\psi_{\vec s}=
-\nabla_{\varphi}\psi_{\vec s}$. Finally, we
express the angular momentum operator from (\ref{deriv}) as
\be
\label{L_3}
-i{\pd\over\pd\varphi} = -{2m_*c\over e\hbar}\,
{\pd H\over\pd B_0} + \pi\xi_0 r^2 + (\mathrm{sgn\,}e)\:
{2\pi\over\Phi_0}\, r A_{1\varphi}(r) \,,
\ee
which allows us to cast the sought angular component into the form
$$
V_{\theta}(\rho) = {1\over\rho}
\left[ -{2m_*c\over e\hbar}\, {\pd E_0(\vec s)\over\pd B_0}
+ \pi\xi_0 \langle\psi_{\vec s}\,|\, r^2\,|\, \psi_{\vec s}\rangle
 + (\mathrm{sgn\,}e)\:
{2\pi\over\Phi_0}\,\langle\psi_{\vec s}\,|\, r A_{1\varphi}(r)\,|\,
\psi_{\vec s}\rangle \right]
$$
\be
\label{angular F}
={1\over\rho} \left\lbrack -{m_*\over
\pi\hbar^2}\, {\pd E_0(\vec s)\over\pd \xi_0} + \pi\xi_0
\langle\psi_{\vec s}\,|\, r^2 \,|\,\psi_{\vec s}\rangle +\,
(\mathrm{sgn\,}e)\: {2\pi\over\Phi_0}\,\langle\psi_{\vec s}\,|\, r
A_{1\varphi}(r)\,|\, \psi_{\vec s}\rangle \right\rbrack\!.
\ee
We stress that in view of (\ref{Green}) $\,V_{\theta}$ is
independent of $\theta$.


\setcounter{equation}{0} \section{Examples}

Let us now illustrate the Berry phase behaviour on several
examples.

\subsection{A homogeneous field} \label{hom}

Suppose that the magnetic field is uniform, \ie, $\vec A_1=0$.
Since the Green and Krein functions are explicitly known in
this case, it is convenient to evaluate the Berry potential
directly from the relation (\ref{berry pot}). It follows from
(\ref{hom G}) that $\psi_{\vec s}$ is of the form
\be
\label{hom psi}
\psi_{\vec s}\,(\vec r\,) = \exp \left\lbrack -\pi
i\xi_0 (\vec r\wedge\vec s\,)\right\rbrack f(|\vec r- \vec s\,|)\,,
\ee
and therefore
\ba
\label{nabla psi}
\nabla_{\theta} \psi_{\vec s}\,(\vec r\,)
&\!=\!& -\pi i\xi_0\, r\rho\, (\cos\varphi\cos\theta
+\sin\varphi\sin\theta)\, \exp \left\lbrack -\pi i\xi_0 (\vec
r\wedge\vec s\,)\right\rbrack f(|\vec r- \vec s\,|) \nonumber \\ &&
+ \exp \left\lbrack -\pi i\xi_0 (\vec r\wedge\vec s\,)\right\rbrack
\nabla_{\theta} f(|\vec r- \vec s\,|)\,.
\ea
Inspecting the explicit form of the function $f$ we see that it is
real-valued and normalized, $\|f\|^2=1$. It follows that
$$
\langle  \psi_{\vec s}\,|\, \exp \left\lbrack -\pi i\xi_0 (\vec
r\wedge\vec s\,)\right\rbrack \nabla_{\theta}\,|\,f \rangle = \langle
f|\, \nabla_{\theta}\,|\, f \rangle = {1\over 2} \nabla_{\theta}
\|f\|^2 = 0\,,
$$
so the sought quantity is given by the first term only,
\ba \label{hom psi2} \lefteqn{ \langle  \psi_{\vec s}\,|\,
\nabla_{\theta}\,|\,\psi_{\vec s} \rangle = -\pi
i\xi_0\,\int_{\R^2} \vec r\cdot \vec s\, |f(|\vec r- \vec s\,|)|^2
\D\vec r} \nonumber \\ && = -\pi i\xi_0\,\int_{\R^2} (\vec
s\,{}^2+ \vec r\cdot \vec s\,) |f(|\vec r\,|)|^2 \D\vec r
\\ && = -\pi i\xi_0 \bigg( \rho^2\! \int_{\R^2} |f(|\vec
r\,|)|^2 \D\vec r  + \rho \int_{\R^2} r(\cos\varphi\cos\theta
+\sin\varphi\sin\theta) |f(|\vec r\,|)|^2 \D\vec r \bigg).
\nonumber \ea
The first integral obviously equals one and the second zero, hence
\be
\label{F theta hom}
V_{\theta}(\rho) = {i\over\rho} \langle
\psi_{\vec s}\,|\, \nabla_{\theta}\,|\, \psi_{\vec s} \rangle =
\pi\xi_0\rho
\ee
and the Berry phase is given by
\be
\label{hom berry} \gamma(C) = 2\pi\xi_0 S\,,
\ee
where $S$ is the area encircled by the loop $C$. We can write it
also as
\be
\label{hom berry2} \gamma(C) = 2\pi\, \mathrm{sgn\,}e\,\frac{\Phi_C}{\Phi_0}
\,,
\ee
where $\Phi_C$ is the full magnetic flux
through the loop. Comparing (\ref{F theta hom}) which corresponds
to $\vec A_1=0$ with the general expression (\ref{angular F})
derived in the previous section, we get in the limit $\rho \to 0$
the relation
\be
\label{hom deriv}
{\pd E_0\over\pd B_0} = {m_*\over 4B_0}\,
\omega_c^2 \langle\psi_{\vec 0}\,|\, r^2\,|\, \psi_{\vec 0}\rangle.
\ee
Let us finish the example with a remark concerning an extension of
the above result to three-dimensional systems. Suppose that the
field is parallel to the $z$-axis and $\vec B(\rho,\varphi,z)=
\vec B(\rho)$ holds in the cylindrical coordinates. Then we have
$V_{\rho}=0,\, V_{\theta}=\pi\xi_0\rho$, and $V_{\zeta}=0$, where
$(\rho,\theta,\zeta)$ are the cylindrical coordinates of the point
$\vec s$. Consequently, the Berry phase along a closed loop $C$ is
again $\gamma(C)=2\pi\Phi_C/\Phi_0$, up to a sign, where $\Phi_C$ is now
the magnetic flux through the projection of $C$ to a
plane perpendicular to the field.

\subsection{A magnetic whisker} \label{whi}

The opposite extreme corresponds to the situation where the
homogeneous component is absent and the field is concentrated into
a flux line (sometimes called Aharonov-Bohm solenoid), \ie, $\vec
B_0=0$ and $\vec A_1= {\eta\Phi_0\over 2\pi r}\, \vec
e_{\varphi}$. Then (\ref{angular F}) yields
\be
\label{F theta whisk}
V_{\theta}(\rho) = - {m_*\over
\pi\hbar^2\rho}\, {\pd E_0\over\pd \xi_0} + {\eta\over\rho}\,
\mathrm{sgn\,}e\,.
\ee
Suppose, in particular that the point perturbation moves along a
circle $C$ of radius $R$ centered at the origin of coordinates. In
that case the Berry phase equals
\be
\label{whisk berry}
\gamma(C) = -\left. {2m_*\over \hbar^2}\, {\pd
E_0\over\pd \xi_0}\right|_{B_0=0} + 2\pi\,(\mathrm{sgn\,}e)\,\eta\,.
\ee
The total flux $\Phi_C$ of the field $\vec B$ through the circle
$C$ is $(\pi R^2\xi_0+\eta)\Phi_0$. Keeping the flux $\Phi$ fixed,
we have ${\pd\over\pd\xi_0}=\pi R^2\Phi_0{\pd\over\pd\Phi_C}$.
Hence
\be
\label{whisk berry2}
\gamma(C) = \left[- {2\pi m_*R^2\Phi_0\over
\hbar^2}\, {\pd E_0\over\pd \Phi_C} + 2\pi(\mathrm{sgn\,}e)
\frac{\Phi_C}{\Phi_0}\right]_{B_0=0}\,.
\ee
Recall that for a particle confined to the loop $C$ the derivative
$\pd E_0/\pd \Phi_C$ equals $-{1\over c}I_0$ where $I_0$ is the
corresponding persistent current. To understand better the meaning
of the Eq.~(\ref{whisk berry2}) we consider the following example.

\subsection{Electron in a ring} \label{ring}

Up to now the confining potential of (\ref{Ham1}) was trivial. The
previous example inspires us to analyze another extreme situation
in which $W$ is a very deep and narrow well. To get a solvable
model we employ the usual idealization and suppose that the
particle is confined to an infinitely thin circular ring $C$ pierced
by the magnetic field. In that case the Hamiltonian $H$ becomes
one-dimensional. Having in mind an electron, $e<0$, we can write $H$ as
\be
\label{ring Ham}
H = {\hbar^2\over 2m_*R^2} \left(
-i{\pd\over\pd\varphi} +\eta \right)^2,
\ee
where $R$ is the ring radius and $\Phi=\eta \Phi_0$ is the total flux
of the field $\vec B$ through the circle; the field profile is
irrelevant here. The Green and Krein function are of the form
\be
\label{ring Green}
G(\varphi,\varphi';E) = {m_*R\over
\pi\hbar^2} \sum_{m=-\infty}^{\infty} {e^{im(\varphi-\varphi')}
\over (m+\eta)^2 - {2m_*R^2\over \hbar^2}E}\,,
\ee
and
\ba
\label{ring Krein}
Q(E;\eta) &\!=\!& {m_*R\over \pi\hbar^2}
\sum_{m=-\infty}^{\infty} \left\lbrack (m+\eta)^2 - {2m_*R^2\over
\hbar^2}E \right\rbrack^{-1} \nonumber \\ &\!=\!& {m_*\over \hbar
\sqrt{m_*E}}\: {\sin {2\pi R\over\hbar}\sqrt{2m_* E} \over \cos
{2\pi R\over\hbar}\sqrt{2m_* E} - \cos 2\pi\eta} \,.
\ea
Consider now a point perturbation of the operator $H$,
\be
\label{ring pert} H_{\theta} = H +
\alpha^{-1}\delta(\varphi-\theta)\,.
\ee
As above the Green function for $H_{\theta}$ is given by the Krein
formula
\be
\label{ring Krein f} G_{\theta}(\varphi,\varphi';E) =
G(\varphi,\varphi';E) - [Q(E)+\alpha]^{-1} G(\varphi,\theta;E)
G(\theta,\varphi';E)\,.
\ee
A solution to the spectral condition
\be
\label{ring spect} Q(E)+\alpha = 0
\ee
exists in each interval $(\tilde E_{\ell}, \tilde E_{\ell+1})$,
where $\{ \tilde E_{\ell}\}_{\ell\ge0}$ is the sequence of ``free"
eigenvalues
\be
\label{ring free ev} \tilde E^{(m)} = {\hbar^2\over 2m_*R^2}\,
(m+\eta)^2
\ee
arranged in the ascending order. In addition, for $\alpha<0$ the
Eq.~(\ref{ring spect}) has a solution also on the halfline
$(-\infty,\tilde E_0)$.

Consider a fixed solution $E_0(\theta)$ of (\ref{ring spect}). It
is clearly independent of $\theta$ and represents a nondegenerate
eigenvalue of $H_{\theta}$ with the eigenfunction
\be
\psi_{\theta}(\varphi) = \left\lbrack {\pd Q\over \pd E} (E_0)
\right\rbrack^{-1/2} G(\varphi,\theta;E_0)
\ee
Let us evaluate the Berry phase when the perturbation site $\theta$ travels
once around the ring. The Berry potential is given by
\be
\label{ring Berry pot}
V(\theta) = i\langle\psi_{\theta} |\,
\nabla_{\theta}\,|\, \psi_{\theta} \rangle\,.
\ee
We express $\psi_{\theta}$ in the form
\be
\label{ring Krein ef}
\psi_{\theta}(\varphi) = {m_*Rc_0 \over
\pi\hbar^2} \sum_{m=-\infty}^{\infty} {e^{im(\varphi-\theta)}
\over (m+\eta)^2 - {2m_*R^2\over \hbar^2}E }
\ee
with $c_0:= \left\lbrack {\pd Q\over \pd E} (E_0)
\right\rbrack^{-1/2}$. Then
\be
\label{ring Berry pot2}
V(\theta) = {2m_*^2R^3c_0^2 \over
\pi\hbar^4} \sum_{m=-\infty}^{\infty} {m \over \left((m+\eta)^2 -
{2m_*R^2\over \hbar^2}E \right)^2} .
\ee
On the other hand,
\be
\label{coef C}
{\pd Q\over \pd E} (E_0) = {2m_*^2R^3 \over
\pi\hbar^4} \sum_{m=-\infty}^{\infty} \left\lbrack (m+\eta)^2 -
{2m_*R^2\over \hbar^2}E \right\rbrack^{-2},
\ee
so
\be
\label{ring Berry pot3}
V(\theta) =\sum_{m=-\infty}^{\infty}
{m \over \left((m+\eta)^2 - {2m_*R^2\over \hbar^2}E \right)^2}
\left\{ \sum_{m=-\infty}^{\infty} \left\lbrack (m+\eta)^2 -
{2m_*R^2\over \hbar^2}E \right\rbrack^{-2} \right\}^{-1}\!.
\ee
Differentiating now $Q$ with respect to $\eta$ we get
\be
\label{Q/Phi}
{\pd Q\over \pd\eta} = -{2m_*R \over
\pi\hbar^2} \sum_{m=-\infty}^{\infty} {m+\eta \over
\left((m+\eta)^2 - {2m_*R^2\over \hbar^2}E \right)^2}
\ee
which yields the identity
\be
\label{ring ident}
\sum_{m=-\infty}^{\infty} {m \over
\left((m+\eta)^2 - {2m_*R^2\over \hbar^2}E \right)^2} =
-{\pi\hbar^2 \over 2m_* R}\, {\pd Q\over \pd\eta}-\!
\sum_{m=-\infty}^{\infty} {\eta \over \left((m+\eta)^2 -
{2m_*R^2\over \hbar^2}E \right)^2}\,.
\ee
In combination with (\ref{coef C}) and (\ref{ring Berry pot3})
this formula gives
\be
\label{ring Berry pot4}
V(\theta) = -{m_*R^2 \over \hbar^2}\, {\pd E_0\over \pd\eta} - \eta
\ee
and the corresponding Berry phase accumulated while $\theta$ moves
once anticlockwise around $C$ is
\be
\label{ring Berry}
\gamma(C) = -{2\pi m_*R^2 \over \hbar^2}\,
{\pd E_0\over \pd\eta} - 2\pi\eta\,.
\ee
Taking into account that the total flux through the ring is
$\Phi_C =\eta\Phi_0$ we see that the obtained expression is fully
analogous to the formula (\ref{whisk berry2}) valid in the whisker
case.

\subsection{A parabolic quantum dot} \label{dot}

As the next example of this section we shall discuss a quantum dot
in a uniform magnetic field $\vec B_0$. To get a solvable
model, we suppose that the confining potential which determines
the dot is parabolic, $W(r)= {1\over 2}m_*\omega_0^2 r^2$. The
frequency $\omega_0$ is related to the effective radius $R$ of the
dot by
\be
\label{eff R}
\zeta = {1\over 2}m_*\omega_0^2 R^2\,,
\ee
where $\zeta$ is the chemical potential of the system \cite{BL}.
The spectrum of $H$ is discrete with the eigenvalues (usually called the
Fock--Darwin levels)
\be
\label{FD levels} E_{mn} = \hbar\omega \left( {|m|+1\over 2} +
n \right) + \hbar\omega_c n\,, \quad m\in\Z, \, n\in\N\,,
\ee
where $\omega:= \sqrt{\omega_c^2+ \omega_0^2}$.
We can employ the known propagator kernel \cite{KC} of the operator $H$,
\ba
\label{dot kern}
K(\vec r,\vec r\,';t) &\!=\!& {m_*\omega \over
4\pi i\hbar \sin {\omega t\over 2}}\, \exp\Bigg\{ {im_*\omega
\over 4\hbar \sin {\omega t\over 2}} \Bigg\lbrack (r^2+r\,'^2) \cos
{\omega t\over 2} \nonumber \\ && - 2\vec r\cdot\vec r\,' \cos
{\omega_c t\over 2} - 2i \vec r\wedge \vec r\,' \sin {\omega_c
t\over 2} \Bigg\rbrack \Bigg\} \,,
\ea
To find an integral representation of the Green functions of
$H$, one has to perform the Wick rotation in (\ref{dot kern}),
\ie, to pass to the imaginary time $t\to -it$. This yields the
heat kernel of $e^{-tH}$; applying the Laplace transformation to
it we get
\ba
\label{dot G kern}
G(\vec r,\vec r\,';t) &\!=\!& {m_*\omega
\over 2\pi\hbar^2}\, \int_0^{\infty} e^{2tE/\hbar} \exp\Bigg\{
-{m_*\omega \over 4\hbar \sinh \omega t} \Bigg\lbrack (r^2+r\,'^2)
\cosh \omega t \nonumber \\ && - 2\vec r\cdot\vec r\,' \cosh
\omega_c t + 2i \vec r\wedge \vec r\,' \sinh\omega_c t \Bigg\rbrack
\Bigg\}\, {\D t\over \sinh\omega t} \,.
\ea
We shall also need the Krein function. It is obtained by the
following trick: we observe that replacing $\omega_c$ at the \rhs
of (\ref{dot G kern}) by $\omega$ we get the Green function of
the Landau Hamiltonian with the cyclotronic frequency $\omega$. We
add and subtract this function at the \rhs, then we subtract the
singularity, ${m_*\over\pi\hbar^2}\, \ln|\vec r-\vec r\,'|^{-1}$,
and pass to the limit $\vec r,\vec r\,'\to\vec s$. In accordance
with (\ref{hom Q}) we obtain
\ba
\label{dot Krein}
Q(E;\vec s\,) &\!=\!& {m_*\omega \over
2\pi\hbar^2}\, \int_0^{\infty} e^{2tE/\hbar} \exp\Bigg\{
-{m_*\omega \over 2\hbar \sinh \omega t} \rho^2 ( \cosh \omega t -
\cosh \omega_c t) - 1 \Bigg\} \nonumber
\\ && \times {\D t\over \sinh\omega t}
-\,{m_*\over 2\pi\hbar^2}\, \left\lbrack \psi\left( {1\over 2} -
{E\over\hbar\omega} \right) +2\gamma -\ln2 -2\ln a
\right\rbrack,\ea
where $a:= \sqrt{\hbar\over m_*\omega}$.

We shall not analyze the last expression generally and restrict
ourselves to showing that if the point-interaction is strong
enough in the sense that $E_0\ll-\hbar\omega$, the confinement
potential has an insignificant effect on the Berry potential only.
To this aim we denote $2E/\hbar=-\varepsilon$ and split the integral $I$
in \rhs of (\ref{dot G kern}) into a sum $I=I_1(\varepsilon)+
I_2(\varepsilon)$ of integrals corresponding to
the intervals $(0,\varepsilon^{-1/2})$ and $(\varepsilon^{-1/2},
\infty)$. It is easy to see that the first integral obeys the
inequality $I_1(\varepsilon)\ge c_1(\vec r,\vec r\,')e^{-\sqrt{\varepsilon}}$
with a constant $c_1$ depending on $\vec r$ and $\vec r\,'$ only. On the
other hand, using an integration by parts we
find that $I_2(\varepsilon)\le c_2(\vec r,\vec r\,')\varepsilon^{-1}
e^{-\sqrt{\varepsilon}}$.
Neglecting for large $|E|$ the second integral, we have
\ba
\label{approx G}
G(\vec r,\vec r\,';t) &\!\simeq\!& {m_*\omega
\over 2\pi\hbar^2}\, \exp\left\{ -i{m_*\omega_c \over 2\hbar} \,
\vec r\wedge \vec r\,'\right\} \nonumber \\ && \times
\int_0^{\varepsilon^{-1/2}} e^{2tE/\hbar} \exp\left\{ -{m_*\omega
\over 4\hbar t} (\vec r-\vec r\,')^2\right\}\, {\D t\over t}\,.
\ea
Since the integral depends on $|\vec r-\vec r\,'|^2$ only, we can
repeat the considerations of Section~\ref{hom} obtaining thus
\be
\label{parab V}
V_{\theta} = \pi\xi_0\rho +\OO(|E|_0^{-1})\,. \ee

\subsection{A zero-range quantum dot} \label{zr dot}

The results of the previous section may be better understood by
considering the zero-range limit of the confinement potential
$W(r)$ of the dot. Specifically, let
\be
\label{pseudoW}
W(r)=\mu_0\delta(\vec r\,)\left(1-(\ln r)
\vec r \,\nabla_{\vec r}\right),
\ee
in accordance with Eq.~(\ref{pseudo}). Then the Green function of
$H$ has the form (see (\ref{krein})):
\be
\label{kreinW}
G(\vec r,\vec r\,';E) = G_0(\vec r,\vec r\,';E) -
[Q_0(E)+\alpha_0]^{-1} G_0(\vec r,0;E)G_0(0,\vec r\,';E)\,,
\ee
where $\alpha_0=\mu_0^{-1}$, $G_0(\vec r,\vec r\,';E)$ is given by
the \rhs of (\ref{hom G}) and $Q_0(E)$ is equal to the expression
at \rhs of Eq.~(\ref{hom Q}). Hence
\be
\label{zr Q}
Q(\vec s\,;E)=
Q_0(E)-[Q_0(E)+\alpha_0]^{-1} G_0^2(\vec s,0;E)\,,
\ee
and
\ba \psi_{\vec s}(\vec r\,) &\!=\!& \left(\frac{\pd Q(\vec
s\,;E)}{\pd E}\right)^{-1/2} \Big\lbrack G_0(\vec r,\vec s\,;E)
\nonumber \\ && -(Q_0(E)+\alpha_0)^{-1} G_0(\vec r,0;E)G_0(0,\vec
s\,;E)\Big\rbrack\,. \label{psi zr} \ea
Since $G_0(\vec s,0;E)$ is independent of $\theta$, we have
$$ \nabla_{\theta}\,\psi_{\vec s}= \left(\frac{\pd Q(\vec
s\,;E)}{\pd E}\right)^{-1/2}\nabla_{\theta}\, G_0(\vec r,\vec
s\,;E)\,. $$
Using now the results of Section 4.1, we obtain
\ba \lefteqn{\langle \psi_{\vec s}\,|\, \nabla_{\theta}\,|\,
\psi_{\vec s} \rangle = \left(\frac{\pd Q(\vec s\,;E)}{\pd
E}\right)^{-1}\Bigg[ \left(\frac{\pd Q_0(E)}{\pd E}\right)(-\pi
i\xi_0\rho^2)} \nonumber \\ && \label{bra zr}
-(Q_0(E)+\alpha_0)^{-1}G_0(0,\vec s\,;E) \langle G_0(\vec
r,0;E)\,|\, \nabla_{\theta}\,|\, G_0(\vec r,\vec s\,;E)
\rangle\,\Bigg]. \phantom{AAA} \ea
It is clear that
\be
\label{inv zr}
\langle G_0(\vec r,0;E)\,|\, \nabla_{\theta}\,|\,
G_0(\vec r,\vec s\,;E) \rangle =\nabla_{\theta}\,
\langle G_0(\vec r,0;E)\,|\, G_0(\vec r,\vec s\,;E) \rangle\,.
\ee
On the other hand the scalar product
$\langle G_0(\vec r,0;E)\,|\, G_0(\vec r,\vec s\,;E) \rangle$
has the form
\be
\label{int zr}
\langle G_0(\vec r,0;E)\,|\, G_0(\vec r,\vec s\,;E) \rangle=
\int\limits_{\R^2}\exp (-\pi i\xi_0 \vec r\wedge\vec s\,)\,
f(|\vec r\,|)\,g(|\vec r-\vec s\,|)\, \D\vec r\,,
\ee
and therefore it is invariant with respect to rotations of the
vector $\vec s$ around the origin. Indeed, let $T$ be such a
rotation, then
\ba \lefteqn{\int\limits_{\R^2}\exp (-\pi i\xi_0 \vec r\wedge T
\vec s\,)\, f(|\vec r\,|)\,g(|\vec r-T\vec s\,|)\, \D\vec r\,}
\nonumber \\ && = \int\limits_{\R^2}\exp (-\pi i\xi_0 T\vec
r\wedge T\vec s\,)\, f(|T\vec r\,|)\,g(|T\vec r-T\vec s\,|)\,
\D\vec r\,\nonumber \\ && = \int\limits_{\R^2}\exp (-\pi i\xi_0
\vec r\wedge\vec s\,)\, f(|\vec r\,|)\,g(|\vec r-\vec s\,|)\,
\D\vec r\,. \nonumber \ea
As a result, Eqs.~(\ref{bra zr}) and (\ref{inv zr}) lead to the
following expression for the non-zero component of the Berry
potential:
\be
\label{Ber zr}
V_\theta(\rho)=
\left(\frac{\pd Q(\vec s\,;E)}{\pd E}\right)^{-1}
\left(\frac{\pd Q_0(E)}{\pd E}\right)(\pi \xi_0\rho)\,.
\ee
Using the asymptotics $$ Q_0(E)=\OO(\ln|E|), \quad G_0(\vec
s\/,0;E)=\OO(|E|^{-1})\quad {\rm as}\quad E\to -\infty\,, $$ we
see from Eqs.~(\ref{zr Q}) and (\ref{Ber zr}) that in a deep
zero-range well, in the sense that $E_0\ll-\hbar\omega_c$, we get
\be
\label{approx zr}
V_{\theta} = \pi\xi_0\rho +\OO(|E|_0^{-2})\,
\ee
in accordance with the result (\ref{parab V}) of the previous
example.


\setcounter{equation}{0} \section{Wilczek--Zee phase}

It was essential in the above considerations that the energy level
in question was nondegenerate. In the opposite case the behavior
of the system with respect to a moving perturbation is more
complex, the degenerate levels may form different linear
combinations and the change includes more than a simple phase
factor. Nevertheless, the effect is usually labeled as the
Wilczek--Zee phase \cite{WZ}.

In magnetic systems with a homogeneous field a prime example of a
degenerate eigenvalue are the Landau levels which constitute the
spectrum of the unperturbed operator (\ref{Ham1}) with $\vec
B_1=0$ and $W=0$; they are
\be
\label{LL}
\varepsilon_{\ell} = \left( \ell+{1\over 2} \right)
\hbar\omega_c\,, \quad \ell=0,1,2,\dots\,.
\ee
In this section we will briefly discuss how the corresponding
eigenfunctions behave under the influence of a moving point
interaction.

Let us first observe that the perturbation preserves the Landau
levels as infinitely degenerate eigenvalues. Let $L_{\ell}$ be the
eigenspace of $H$ referring to an eigenvalue $\varepsilon_{\ell}$.
It is straightforward to see that the eigenspace of $H_{\vec s}$
corresponding to the same eigenvalue has the following form
\be
\label{Land esp} L_{\ell}(\vec s\,) = \{ \psi\in L_{\ell}:\:
\psi(\vec s\,)=0\, \}\,.
\ee
Since $L_{\ell}$ is invariant with respect to translations of the
eigenfunctions, it is possible to select an orthonormal basis
$\psi^{(\ell)}_1(\vec s\,), \psi^{(\ell)}_2(\vec s\,), \dots,
\psi^{(\ell)}_n(\vec s\,), \dots$ in
$L_{\ell}(\vec s\,)$ which depends smoothly on the point $\vec
s\in\R^2$. We suppose that $\vec s$ is adiabatically moving along
a smooth closed contour, $\vec s=\vec s\,(t), t\in[0,1]$, and that
at the initial moment $t=0$ the systems is in a state $\psi^{(\ell)}_m(\vec
s\,(0))$. Then the state $\psi(t)$ at an instant $t$ is given by the
formula
\be
\label{WZ}
\psi(t) = e^{\varepsilon_{\ell}t/i\hbar} \sum_n
U^{(\ell)}_{nm}(t) \psi^{(\ell)}_n(\vec s\,(t))\,,
\ee
where $\big(U^{(\ell)}_{nm}(t)\big)$ is a unitary matrix generalizing the
Berry phase factor $e^{i\gamma(t)}$ (see \cite{WZ}).
The role of Berry potential is played by the infinite self-adjoint matrix
\be
\label{Lambda}
V^{(\ell)}_{mn}(\vec s\,) = i\langle\psi^{(\ell)}_m(\vec s\,)\,|\,
\nabla_{\vec s}\,|\, \psi^{(\ell)}_n(\vec s\,)\rangle \,,
\ee
which is related to $U^{(\ell)}(t)\equiv U(t)$ by
\be
\label{LambdaU}
\left(U^{-1}(t)\dot U(t)\right)_{mn} = iV^{(\ell)}_{nm}(\vec s\,(t))\,.
\ee
The solution to the equation
(\ref{LambdaU}) along the curve $C:\vec s= \vec s\,(t)$ is at that
given by the path integral (the Wilson loop)
\be
\label{path i}
U(C) = \mathcal{P} \exp\left( i\oint_C V(\vec s\,)\,
\D \vec s \right) \,,
\ee
where $\mathcal{P}$ indicates a time-ordered exponential.

The Wilczek--Zee theory has the following differential-geometric
interpretation \cite{VDDMS}. Consider the trivial vector bundle
$\EE_\ell=\R^2\times L_\ell$, then $\FF_\ell=\bigcup\left\{\{\vec
s\,\}\times L_\ell(\vec s\,): \,\vec s\in\R^2\right\}$ is a
subbundle of $\EE_\ell$ with the infinite-dimensional typical
fiber $\ell^2$. Denote by $\mathfrak{lu}(\infty)$ the Lie algebra
of the unitary group of $\ell^2$ (the Lie algebra of
skew-Hermitian infinite-dimensional matrices). Then it is
convenient to regard $-iV_{mn}(\vec s\,)$ as coefficients of the
differential form $\omega=\omega_k \D x^k$ assuming values in
$\mathfrak{lu}(\infty)$:
\be
\label{omeg}
\omega_k =\langle\psi^{(\ell)}_m(\vec s\,)\,|\,
\nabla_{x^k}\,|\, \psi^{(\ell)}_n(\vec s\,)\rangle \,\qquad
\vec s=(x^1,x^2)\,.
\ee
This form is a connection form in the bundle $\FF_\ell$, and the
operators $U(C)$ are the holonomy operators in the principal
$\UU(\infty)$-bundle associated with $\FF_\ell$. According to the
Ambrose--Singer theorem \cite{KN}, the curvature form $\Omega$,
$\Omega=\D \omega+\omega\wedge\omega$ determines completely the
operators $U(C)$ (the tensor $F_{jk}=i\Omega_{jk}$ is the strength
of the gauge potential $V_k$). Notice that there is an explicit
formula (analogous to the Stokes formula) which expresses the \rhs
of Eq.~(\ref{path i}) in terms of the coefficients of $\Omega$
\cite{Me}; nevertheless, it is difficult to use this formula when
the components of $\omega$ are not commuting (which is the case
for the matrices (\ref{omeg})). However, we can gain some insight
into the behaviour of the Wilczek--Zee phase considering
infinitely small loops. In particular, for such a loop $C$
encircling a point $\vec s_0$ the holonomy operator is given by an
ordinary exponential
\be
\label{stok}
U(C) = \exp\left(\Omega_{12}(\vec s_0)S \right) \,,
\ee
where $S$ is the area encircled by the loop $S$.

In the following we shall consider for simplicity the lowest
Landau level $\varepsilon_0$ and drop the superscript $0$ for the
notations. Normalized eigenfunctions of the ground state $L_0$ may
be chosen in the form \cite{LL}
\be
\label{ground LL}
\Psi_m(r,\varphi) = \left( |\xi_0|\over 2^m m!
\right)^{1/2} e^{\sigma im\varphi} e^{-r^2/4a_0^2} \left( r\over a_0
\right)^m \,, \quad m\ge 0\,,
\ee
where $\sigma={\rm sgn}\,\xi_0$. The
integral kernel $P_0(\vec r, \vec r\,')$ of the projection operator
onto the subspace $L_0$ equals \cite{Ge}
\be
\label{kernel_0} P_0(\vec r, \vec r\,') = |\xi_0| e^{-\pi i
\xi_0\vec r\wedge\vec r\,'} e^{-(\vec r-\vec r\,')^2/4a_0^2}\, . \ee
The condition $\psi(\vec s\,)=0$ can be then written as
\be
\label{zero node}
\int_{\R^2} P_0(\vec s, \vec r\,) \psi(\vec r\,)\,\D\vec r = 0\,,
\ee
and a comparison with (\ref{ground LL}) shows that this is
equivalent to
\be
\label{zero node2}
\langle [\vec s,\zeta]\Psi_0\,|\psi\rangle =
0\,,
\ee
where $[\vec s,\zeta]$ with $\vec s\in\R^2$ and $\zeta\in\SSS^1$
denotes the operator of magnetic translation \cite{Za1} which acts
on $f\in L^2(\R^2)$ as
\be
\label{mg transl}
[\vec s,\zeta]f(\vec r\,) = \zeta\, \exp\left(-\pi
i \xi_0 \vec r\wedge \vec s\,) \right) f(\vec r-\vec s\,)\,. \ee
This shows that one can choose the family of the functions
\be
\label{on bas}
\psi_m(\vec s) = [\vec s,1]\Psi_m\,,\quad
k=1,2,\dots
\ee
for orthonormal basis in $L_0(\vec s\,)$. Let us calculate the
corresponding matrix elements $V_{mn}(\vec s\,)$. It is convenient
to perform the calculation in the Cartesian coordinates. Let $\vec
r=(x,y)$, $\vec s=(x',y')$; then
\be
\label{psi(m)}
\psi_m(\vec s\,)(\vec r\,)= \exp\left(-\pi i \xi_0
(xy'-x'y) \right) \Psi_m(x-x',\,y-y')\,.
\ee
Writing $\Psi_m$ as
\be
\label{ground LL2}
\Psi_m(x,y) = \left( |\xi_0|\over 2^m m!
\right)^{1/2} e^{-(x^2+y^2)/4a_0^2} \left( x+\sigma iy\over a_0 \right)^m
\ee
we find
\be
\label{Psider}
{\pd\Psi_m\over \pd x} = -{x\over 2a_0^2}\Psi_m +
{1\over a_0} \sqrt{m\over 2} \Psi_{m-1}\,, \quad {\pd\Psi_m\over
\pd y} = -{y\over 2a_0^2}\Psi_m + {\sigma i\over a_0} \sqrt{m\over 2}
\Psi_{m-1}\,.
\ee
Now we obtain from Eq.~(\ref{Psider}):
$$ {\pd\psi_m\over \pd x'}(\vec s)(x,y) = \pi i\xi_0y\psi_m(\vec
s\,) (x,y)+\exp\left(-\pi i \xi_0 (xy'-x'y) \right)\times $$
\be
\label{psiderx}
\left[\frac{x-x'}{2a_0^2}\Psi_m(x-x',\,y-y')-\frac{1}{a_0}\sqrt{
\frac{m}{2}}\Psi_{m-1}(x-x',\,y-y')\right]\,,
\ee

\medskip

$$ {\pd\psi_m\over \pd y'}(\vec s)(x,y) = -\pi i\xi_0x\psi_m(\vec
s) (x,y)+\exp\left(-\pi i \xi_0 (xy'-x'y) \right)\times $$
\be
\label{psidery}
\left[\frac{y-y'}{2a_0^2}\Psi_m(x-x',\,y-y')-\frac{\sigma i}{a_0}\sqrt{
\frac{m}{2}}\Psi_{m-1}(x-x',\,y-y')\right]\,.
\ee
Hence
$$ \langle\psi_n\,|\nabla_{x'}\,|\psi_m\rangle=\pi
i\xi_0y'\delta_{mn}+ \pi i \xi_0\langle\Psi_n\,|y|\Psi_m\rangle+
$$
\be
\label{matrx}
\frac{1}{2a_0^2}\langle\Psi_n\,|x|\Psi_m\rangle-\frac{1}{a_0}
\sqrt{\frac{m}{2}}\delta_{n,m-1}\,, \ee
and
$$ \langle\psi_n\,|\nabla_{y'}\,|\psi_m\rangle=-\pi
i\xi_0x'\delta_{mn}- \pi i \xi_0\langle\Psi_n\,|x|\Psi_m\rangle+
$$
\be
\label{matry}
\frac{1}{2a_0^2}\langle\Psi_n\,|y|\Psi_m\rangle-\frac{\sigma
i}{a_0} \sqrt{\frac{m}{2}}\delta_{n,m-1}\,. \ee
To find the matrix elements $\langle\Psi_n\,|x|\Psi_m\rangle$ and
$\langle\Psi_n\,|y|\Psi_m\rangle$ we make use of the following
observations: the matrices
$\left(i\langle\psi_n\,|\nabla_{x'}\,|\psi_m\rangle\right)$ and
$\left(i\langle\psi_n\,|\nabla_{y'}\,|\psi_m\rangle\right)$ are
Hermitean, and at the same time, the numbers
$\langle\Psi_n\,|x|\Psi_m\rangle$ and
$\langle\Psi_n\,|y|\Psi_m\rangle$ are real. Taking these facts
into account, we get from Eqs.~(\ref{matrx}) and (\ref{matry})
\be
\label{Her1} \frac{i}{2a_0^2}\langle\Psi_n\,|x|\Psi_m\rangle-
\frac{i}{a_0}\sqrt{\frac{m}{2}}\delta_{n,m-1}=
-\frac{i}{2a_0^2}\langle\Psi_n\,|x|\Psi_m\rangle+
\frac{i}{a_0}\sqrt{\frac{n}{2}}\delta_{m,n-1}\,, \ee

\medskip

\be
\label{Her2} \frac{i}{2a_0^2}\langle\Psi_n\,|y|\Psi_m\rangle+
\frac{\sigma}{a_0}\sqrt{\frac{m}{2}}\delta_{n,m-1}=
-\frac{i}{2a_0^2}\langle\Psi_n\,|y|\Psi_m\rangle+
\frac{\sigma}{a_0}\sqrt{\frac{n}{2}}\delta_{m,n-1}\,. \ee
Thus
\be
\label{Matrx}
\langle\Psi_n\,|x|\Psi_m\rangle=\frac{a_0}{\sqrt2}\left(
\sqrt{n}\delta_{m,n-1}+\sqrt{m}\delta_{n,m-1}\right)\,, \ee

\medskip

\be
\label{Matry}
\langle\Psi_n\,|y|\Psi_m\rangle=\frac{\sigma ia_0}{\sqrt2}\left(
\sqrt{m}\delta_{n,m-1}-\sqrt{n}\delta_{m,n-1}\right)\,. \ee
Since $|\xi_0|^{-1}=2\pi a_0^2$ we have finally
\be
\label{finx}
\langle\psi_n\,|\nabla_{x'}\,|\psi_m\rangle=\pi i\xi_0y'\delta_{mn}
+\frac{1}{\sqrt{2} a_0}\left(\sqrt{n}\delta_{m,n-1}-
\sqrt{m}\delta_{n,m-1}\right)\,,
\ee
\be
\label{finy}
\langle\psi_n\,|\nabla_{y'}\,|\psi_m\rangle=-\pi i\xi_0x'\delta_{mn}
-\frac{\sigma i}{\sqrt{2} a_0}\left(\sqrt{n}\delta_{m,n-1}+
\sqrt{m}\delta_{n,m-1}\right)\,.
\ee
Because the matrices
$\left(\langle\psi_n\,|\nabla_{x'}\,|\psi_m\rangle\right)$ and
$\left(\langle\psi_n\,|\nabla_{y'}\,|\psi_m\rangle\right)$ do not
commute, it is not easy to calculate the path integral (\ref{path
i}), and we turn to Eq.~(\ref{stok}) to gain some insight into the
behaviour of the Wilczek--Zee phase. For this purpose let us
calculate the curvature form $\Omega$. Since $\Omega_{jk}$ is
skew-symmetric (w.r.t. the indices $jk$) it is enough to find the
component $\Omega_{12}$. It is clear from Eqs.~(\ref{finx}) and
(\ref{finy}) that $\D\omega= -\pi i\xi_0\delta_{mn}\D x^1\wedge \D
x^2$. Since the first terms in Eqs.~(\ref{finx}) and (\ref{finy})
are scalar matrices, in order to find $\omega\wedge\omega$, we
must to calculate the commutator of the matrices $(\sqrt{2}
a_0)^{-1}\left(\sqrt{n}\delta_{m,n-1}-\!
\sqrt{m}\delta_{n,m-1}\right)$ and $-\sigma i(\sqrt{2}
a_0)^{-1}\left(\sqrt{n}\delta_{m,n-1}+\!
\sqrt{m}\delta_{n,m-1}\right)$ only. As a result we obtain that
\be
\label{curv}
\Omega=2\pi i \xi_0 \delta_{1m}\delta_{1n}\D x^1\wedge \D x^2\,.
\ee
Therefore, for an infinitely small loop $C$ in the plane $\R^2$
the operator $U(C)$ has the diagonal matrix
\be
\label{WZ U}
U_{mn}(C)=
{\rm diag}\left(\exp(-2\pi i\xi_0 S),\,1,\,1\,,\ldots\,,1\,,
\ldots\,\right)\,,
\ee
where $S$ is the area encircled by the loop $C$. Hence during an
adiabatic evolution along the loop $C$ the state $\psi^{(0)}_1$
with the angular momentum $m=1$ is modified by the Berry-like
factor $\exp(-2\pi i\xi_0S)$; the states with the others angular
momenta $m=2,3,\ldots$ remain unchanged. This behaviour of the
Wilczek--Zee phase is similar to the spectral behaviour of the
Aharonov--Bohm Hamiltonian with an infinitely thin solenoid, which
is described in analogy with a delta-perturbed Hamiltonian by a
self-adjoint extension of a symmetric operator. Namely, the
infinitely thin Aharonov--Bohm solenoid perturbs only two states
with neighbour angular momenta (see, \eg, \cite{BV}). Similarly,
in the case of the Wilczek--Zee phase the point potential changes
two states with neighbour angular momenta: $m=0$ and $m=1$. The
opposite signs in (\ref{hom berry}) and (\ref{WZ U}) can be
interpreted as a ``topological charge conservation". More
specifically, the mappings $\R^2\ni\vec s\mapsto [\vec
s,1]\Psi_m$, $m=0,1,\ldots$ form a basis section of the vector
bundle $\EE_0$. Formula (\ref{Lambda}) with $m,n\ge0$ defines a
connection in this bundle, and it is easy to show that this
connection is flat (\ie,  its curvature vanishes). Thus in
accordance with the Ambrose--Singer theorem, all the Wilson loops
(\ref{path i}) are identity operators, i.e., the "Berry phase" for
this connection is equal to zero. Adding the point potentials of
the same strength $\alpha$ to each point $\vec s\in\R^2$ we split
the bundle $\EE_0$ into a sum of the line bundle $\LL_0$ of the
eigenfunctions in the zero-range well and the bundle $\FF_0$ of
the eigenfunctions remaining on the zeroth Landau level:
$\EE_0=\LL_0\oplus\FF_0$. Equations (\ref{hom berry}) and (\ref{WZ
U}) show that the sum of Berry phases related to the summands is
still zero. This effect is similar to the Berry phase conservation
in the Born--Openheimer problem \cite{Za2}. On the other hand, we
have here an analogy with the Novikov formula for the Chern
numbers of a sum of vector bundles of magneto-Bloch functions
\cite{Nov}. In physical terms the Novikov formula states that the
quantized Hall conductivity of a Bloch--Landau band is the sum of
conductivities of the all magnetic subbands of this band. It
remains to note that the mentioned Chern numbers are integrals of
the curvature form.


\subsection*{Acknowledgment}

The research has been partially supported by GAAS and Czech
Ministry of Education under the contracts 1048801 and ME099.


\end{document}